\documentstyle[aps,multicol,epsf]{revtex}

\begin{document}
\title{\bf Monte Carlo study of the magnetic critical properties \\
           of the two-dimensional Ising fluid }

\author{\bf A.L.Ferreira and W.Korneta\thanks
{permanent address : Faculty of Physics, Technical University, 
Malczewskiego 29, 26-600 Radom, Poland } \\
Departamento de Fisica, Universidade de Aveiro, \\
3800 Aveiro, Portugal.}


\maketitle


\begin{abstract}

A two-dimensional fluid of hard spheres each having a spin $\pm 1$ and
interacting via short-range Ising-like interaction is studied near the second 
order phase transition from the paramagnetic gas to the ferromagnetic gas phase. 
Monte Carlo simulation technique and the multiple histogram data analysis were used. 
By measuring the finite-size behaviour of several different thermodynamic quantities,
we were able to locate the transition and estimate values of various static critical 
exponents. The values of exponents $\beta/\nu$ and $\gamma/\nu$ are close to the ones 
for the two-dimensional lattice Ising model.  However, our result for the exponent
$\nu =1.35$ is very different from the one for the Ising universality class.

\end{abstract}



\begin{multicols}{2}
\narrowtext
\section{Introduction} 

Models with coupled translational and spin degrees of freedom attracted recently considerable 
attention because they can describe several phenomena in amorphous ferromagnets \cite{r1}, 
dilute magnetic alloys and dipolar liquids \cite{r2,r3}. Spins in such systems are not localized 
on lattice sites but are able to move. The systems with coupled spin and translational degrees 
of freedom exhibit a rich variety of phase transitions.  Their phase diagrams were determined 
using the mean spherical approximation, the mean-field theory, density functional methods  
and Monte Carlo (MC) simulation techniques \cite{r1,r2,r3,r4}.  However the critical behaviour 
and universality near phase transitions in these systems attracted only little attention.
In ref.\ \cite{r5} the critical properties of the Heisenberg fluid near magnetic order-disorder 
transition were studied by MC simulations.  The obtained critical exponents differ by a small 
but significant amount from the ones for the lattice Heisenberg model. 
The spin fluid systems resemble lattice-based spin models with annealed site dilution. 
The Blume-Capel model is an example of the lattice-based Ising model with an annealed 
site dilution \cite{r6}. The density of annealed sites in this model is, however, 
not fixed but can fluctuate around an average. 

\begin{figure}[tb]
\begin{center}
   \leavevmode
   \epsfxsize=9cm
   \epsfbox{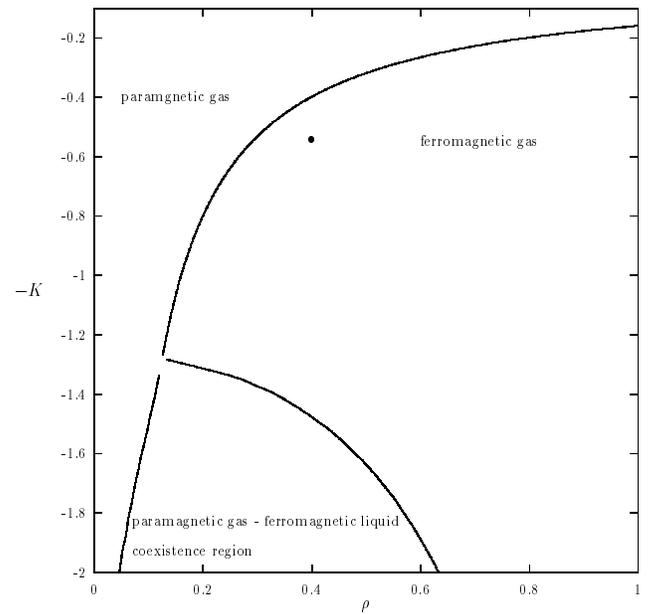}
  \end{center}
\caption{
Mean-field phase diagram for the Ising fluid with the coupling constant $K$ and the 
density $\rho$.  The line separating the paramagnetic and ferromagnetic gas phases 
is given by $K=1/(2 \pi \rho)$.  The point indicates the 
location of the phase transition for $\rho = 0.4$ obtained in this paper from MC 
simulations.}
\label{f1} 
\end{figure}

The phase diagram of the 2d Ising fluid studied in this paper obtained within the mean-field 
approximation described in the ref.\ \cite{r4} is shown in figure \ref{f1}.  For high 
temperatures and low densities there exists critical line 
separating a paramagnetic gas phase from a ferromagnetic gas phase. The critical line finishes
at lower temperatures at the tricritical point.  The similar phase diagram was obtained in
the 2d Blume-Capel model.  Recently MC simulations were used to investigate the tricritical point 
properties of a 2d Ising fluid and 2d Blume-Capel model \cite{r8}. It was shown that both models 
belong to the same tricritical universality class. \\
The aim of this paper is to present the results obtained from MC simulations of the 2d spin fluid 
with short-range Ising-like interactions near the second-order phase transition 
from the paramagnetic gas to the ferromagnetic gas phase far from the tricritical point.  
We performed simulations in systems of different sizes at a constant particle density and
for four different temperatures.  We analyzed our data combining the multiple histogram 
technique \cite{r9} with finite-size scaling (FSS) \cite{r10,r11} to obtain estimates for the 
critical temperature and exponents.  In the Sec.\ II we describe the Ising fluid model and 
technical aspects of the simulations. Determination of critical exponents and the location of 
phase transition are given in Secs. III and IV, while Sec.V summarizes our results.
 \begin{figure}[tb]
\setlength{\epsfxsize}{9cm}
\centerline{\mbox{\epsffile{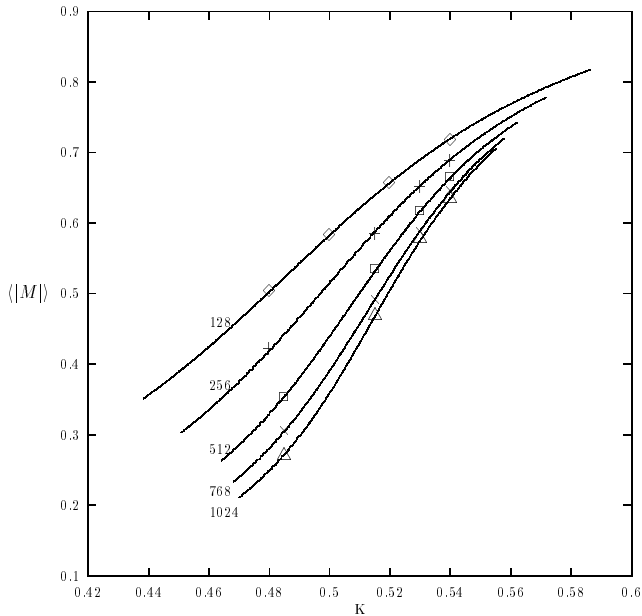}}}

\caption{The magnetization versus the coupling constant $K$ for the five system sizes. 
         The number of particles in the system is indicated.  The points are averages
         over individual simulations.  The curves results from the multiple histogram
         technique.  Error bars are smaller than the symbol size.}
\label{f2} 
\end{figure}
\section{ The model and simulation details }

We consider a system which consists of particles of diameter $\sigma$ in two-dimensional space.  
The internal degrees of freedom of each particle are described by an Ising spin and there is an 
exchange coupling between spins given by the Yukawa interaction.  
The system Hamiltonian is:

\begin{equation} 
\label{e1}
H=\sum_{i,j} \phi(r_{ij}) S_i\cdot S_j 
\end{equation}
where the interaction potential has the following form,

\begin{equation}
\label{e2}
\phi(r) = \left\{ \begin{array}{ll}
        \infty & \mbox{ if $r < \sigma$}  \\ 
        -K \ \frac{\sigma}{r} \ \exp(-(r-\sigma)/\sigma) & \mbox{ if $r \ge \sigma$}
        \end{array}
        \right.
\end{equation}
        
The parameter $K$ is the ratio of the coupling energy to the thermal energy.
In this paper we consider the ferromagnetic case ($K>0$). \\
\begin{figure} [tb]
\setlength{\epsfxsize}{9cm}
\centerline{\mbox{\epsffile{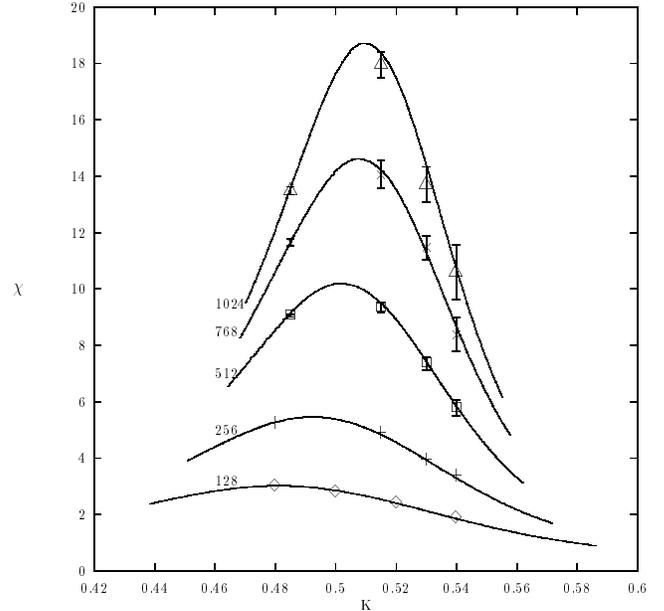}}}

\caption{The magnetic susceptibility versus the coupling constant $K$ for the five system sizes. 
         The number of particles in the system is indicated.  The points are averages
         over individual simulations.  The curves results from the multiple histogram
         technique.  Error bars are omitted when smaller than the symbol size.}
\label{f3} 
\end{figure}
The MC simulations were performed at a constant particle density $\rho=0.4$.  We studied 
five systems with the number of particles $N$ equal to 128, 256, 512, 768 and 1024.  The periodic   
boundary conditions and the minimum image convention were applied during simulations \cite{r12}.
The interaction potential $\phi(r)$ was cut at a distance $6.3246\sigma$.  This value was chosen 
in order to divide the simulation cell into 16 sub-cells for the system with 256 particles.  
In order to speed up simulations we used the method of linked lists of neighbours \cite{r12}.
We applied the same simulation algorithm as it is described in ref.\ \cite{r3}. 
In the present work we have not included any long--range correction
to the cuttof procedure as it was done in \cite{r3}.
The maximum position displacement of particles was chosen in such a way, that the acceptance 
ratio of the trial moves was around 0.5.  The number of 
$MCS/N$ (Monte Carlo steps per particle) discarded at the beginning of the simulation was
larger than $10^{4}$.  For each system size the simulations were performed for four values of
the parameter K.  These values were the following: $K= 0.48, 0.5,0.52$ and $0.54$ for a system 
with 128 particles, $K=0.48,0.515,0.53$ and $0.54$ for a system with 256 particles and 
$K=0.485,0.515,0.53$ and $0.54$ for systems with 512,768 and 1024 particles.  
These values of K were chosen, because the range of temperatures where reliable extrapolation
of the behaviour of physical quantities could be performed included positions of the maxima 
of specific heat and magnetic susceptibility and the position of the phase transition 
in the bulk system.  The data were stored 
at intervals of 10 $MCS/N$ and the total number of updates was $2 \cdot 10^{6}$ $MCS/N$ for
every $K$ value.  

Let's denote by $M$ the magnetization per particle of the system defined as $M=(\sum_{i} S{i})/N$.
We study critical behaviour of the following quantities \cite{r13,r14}: the mean absolute value of 
the magnetization $<|M|>$, the magnetic susceptibility defined as \mbox{$\chi=KN(<M^{2}>-<|M|>^{2})$} 
the fourth-order magnetization cumulant defined as $U=1-<M^{4}>/(3<M^{2}>)$, the mean energy of 
the system $<H>$ and the specific heat $C_{V}=(<H^{2}>-<H>^{2})/N$, where $<...>$ denotes
canonical ensemble average. We also consider the quantities
like the derivatives  $\partial \ln <|M|> / \partial K$ and $\partial \ln <M^{2}> / \partial K$.
In the Heisenberg fluid \cite{r5} and 3d lattice Ising model \cite{r14} these derivatives were
used to extract the critical exponent $\nu$.  \\
\begin{figure} [tb]
\setlength{\epsfxsize}{9cm}
\centerline{\mbox{\epsffile{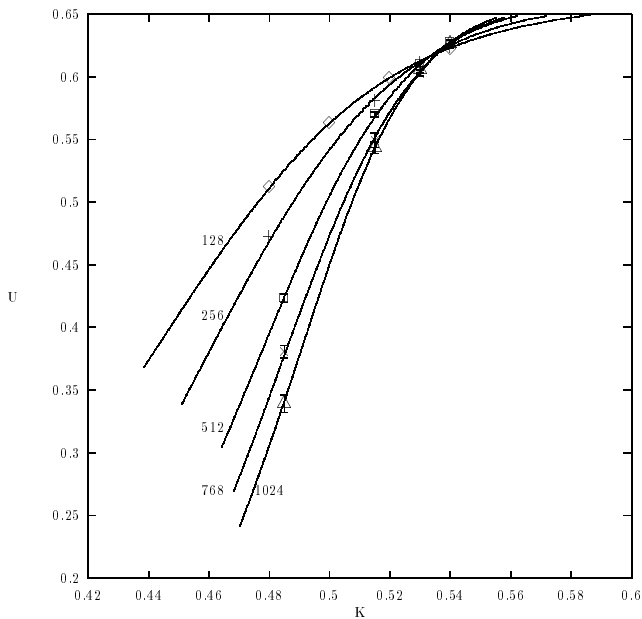}}}

\caption{The fourth-order magnetization cumulant versus the coupling constant $K$ for the 
         five system sizes. 
         The number of particles in the system is indicated.  The points are averages
         over individual simulations.  The curves results from the multiple histogram
         technique.  Error bars are omitted when smaller than the symbol size.}
\label{f4} 
\end{figure}
The thermodynamic properties of a system, in a wide temperature range, can be obtained 
by performing several MC simulations at different temperatures and combining them by 
the application of the multiple histogram technique \cite{r9}. This technique allows reliable 
extrapolation of MC results to the values of K where the interesting positions of the maxima
shown by some of the quantities defined above are located.  In figures \ref{f2}, \ref{f3} and
\ref{f4} we show the dependence of quantities $<|M|>$, $\chi$ and $U$ on K. 
The results obtained directly from MC simulations
performed at selected K values are also shown in these figures by points.  
The error bars were obtained by calculating block averages of $10^{5} MCS/N$ data points 
and computing the standard deviation from these block averages.  One can notice that the 
multiple histogram extrapolations are within calculated error bars.
  
\begin{figure} [htb]
\setlength{\epsfxsize}{9cm}
\centerline{\mbox{\epsffile{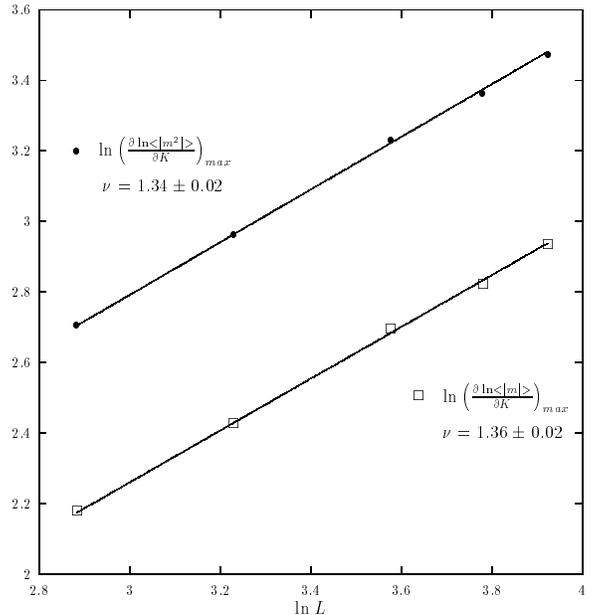}}}

\caption{The plot of the maxima of  $\frac{\partial \ln <|M|>}{ \partial K}$ 
         and $\frac{\partial \ln <M^{2}>}{\partial K}$ versus the linear system size $L$.
         Error bars are smaller than the symbol size. 
         The straight lines are fits to the data.  Their slope is $1 / \nu$.  
         The obtained values of the exponent $\nu$ are indicated.}   
\label{f5} 
\end{figure}
\section{ Estimates of the exponent $\nu$ and the location of the phase transition 
          in the bulk system}

The critical exponent $\nu$ characterizing the divergence of the correlation length 
near the second-order phase transition \cite{r11} can be extracted by considering the scaling 
behaviour of the derivatives $\partial \ln <|M|> / \partial K$ and 
$\partial \ln <M^{2}> / \partial K$ \cite{r5,r14}.  
These derivatives can easily be computed using the following identity: 

\begin{equation}
\label{e3}
\frac{\partial <|M^{n}|>}{\partial K} = - \frac{1}{K} \left( <|M^{n}|H>-<|M^{n}|><H> \right)
\end{equation}

Let's denote by $L$ the length of one side of the simulation box.  In our case $L=\sqrt{N/\rho}$.
The dependence of these derivatives on $K$ has the maximum which should scale with the system 
size as $L^{1/\nu}$.  This method of estimation of the exponent $\nu$ is very convenient, 
because it can be done without any consideration of the critical coupling $K_{c}$ in the bulk 
system.  We show in the figure \ref{f5} maximum values of  $\partial \ln <|M|> / \partial K$ 
and $\partial \ln <M^{2}> / \partial K$ in systems of different sizes together with the fitted 
straight lines. The goodness of fits $Q$ \cite{r15} are $Q=0.36$ and $Q=0.32$ for the derivative 
of $\ln <|M|>$ and $ln <M^{2}>$ respectively.  The slope of these lines provides estimates for 
$\/\nu$, and we obtained $\nu = 1.36 \pm 0.02$ for the maxima of $\partial \ln <|M|> / \partial K$ 
and $\nu = 1.34 \pm 0.02$ for the maxima of $\partial \ln <M^{2}> / \partial K$.  
These values are higher than the value $\nu =1$ in the two-dimensional lattice Ising model.
\begin{figure}[tb]
\setlength{\epsfxsize}{9cm}
\centerline{\mbox{\epsffile{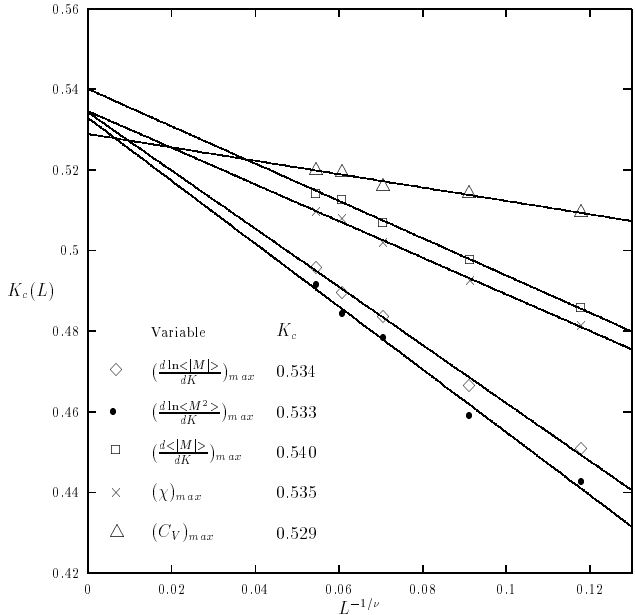}}}

\caption{Size dependence of the location of maxima $K_{c}(L)$ for several quantities indicated 
         in the figure in finite systems containing 128, 256, 512, 768 and 1024 particles.
         The straight lines are fits to Eq.\ \protect{\ref{e4}} with the exponent $\nu = 1.35$.  
         The estimated values of the critical coupling $K_{c}$ in the bulk system obtained 
         from these fits are shown for each quantity.}
\label{f6} 
\end{figure}
The critical coupling $K_{c}$ in the bulk system is usually determined using the Binder 
fourth-order magnetization cumulant crossing technique\cite{r5,r11,r14}.  
Finite size scaling predicts that for sufficiently large systems if we plot $U$ versus $K$ 
for different choices of $L$, these curves should have a unique intersection point $U^{*}$. 
The value of $K$ where this occurs is the value of the critical coupling $K_{c}$.
This value is not biased by any assumptions about critical exponents.
For smaller systems there are corrections to FSS and the intersection point between any two 
curves $U$ vs.\ $K$ corresponding to systems with side lengths $L$ and $L'$  depends on
$L$ and $L'$ \cite{r16}.  In Table I we give the coordinates of intersection points for different 
pairs of systems.
The values in the table have statistical errors larger than the expected correction terms to FSS.
Because of this and a small number of systems studied, we were not able to extract $K_{c}$
by the extrapolation procedures given in ref. \cite{r14}.The critical coupling $K_{c}$ and the 
intersection value $U^*$ we calculated as the average from the values in the table. We obtained
$K_{c}=0.535\pm 0.002$ and $U^{*}=0.618 \pm 0.003$.  We have excluded from the average the last three rows
of the table because for systems with small difference in sizes even a small shift in the
cumulant lines can produce a considerable error in the coordinates of intersection points.
The estimated common value of the cumulant 
is only slightly larger than the value $U^{*}=0.611 \pm 0.001$ obtained for the two-dimensional
lattice Ising model \cite{r17}.

The value of $K_{c}$ can also be determined from the size-dependent shifting of the peak of 
different thermodynamic quantities. In finite systems the quantities like e.g. 
the specific heat $C_{V}$, the magnetic susceptibility $\chi$,  $\partial <|M|> / \partial K$ ,
$\partial \ln <|M|> / \partial K$ and $\partial \ln <M^{2}> / \partial K$, have  maxima as a
function of $K$\cite{r10,r11,r14}. 
 The location of the maximum $K_{c}(L,A)$ depends both on the system size $L$
and on the quantity $A$ considered.  FSS predicts the following dependence of $K_{c}(L,A)$ on
the system size \cite{r10,r14}:

\begin{equation}
\label{e4}
K_{c}(L,A)=K_{c} +aL^{-1 / \nu}
\end{equation}
with the omitted corrections to FSS.  The constant $a$ depends in magnitude and sign on
the particular quantity considered.  In order to determine $K_{c}$ from this equation it is
necessary to have both an accurate estimate of the exponent $\nu$ and accurate values of
$K_{c}(L,A)$.  In figure \ref{f6} we plot estimates of $K_{c}(L,A)$ for different quantities
as a function of $L^{-1 / \nu}$.  The lines in this figure were obtained by the least square 
fits of the data to Eq.\ \ref{e4} with $\nu=1.35$, the average of previously determined values. 
One can notice 
that values of $K_{c}$ obtained from the fit agree well with the estimated value $K_{c}=0.535$
estimated above from the cumulant intersection points.  The value of the exponent $\nu$ 
different than $1.35$ will lead, of course, to different estimates of $K_{c}$ which are
inconsistent with the value extracted from cumulant intersection points.


\section{Estimates of exponent ratios $\beta / \nu$ and $\gamma / \nu$}

\begin{figure} [htb]
\setlength{\epsfxsize}{9cm}
\centerline{\mbox{\epsffile{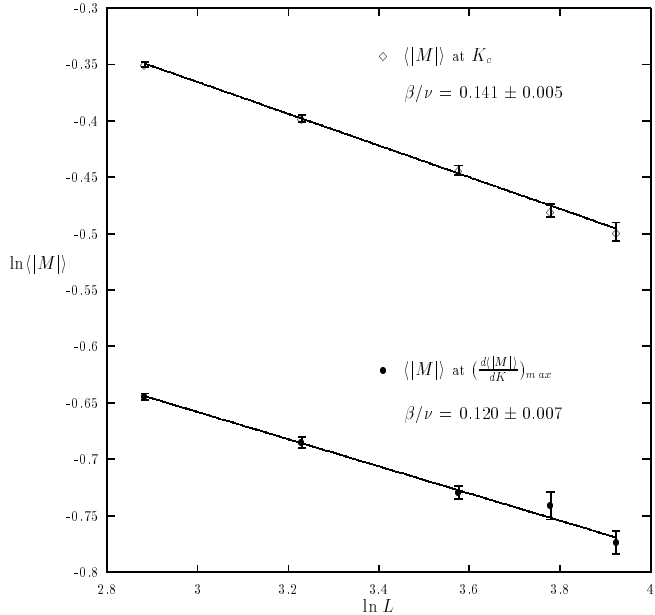}}}

\caption{The plot of the magnetization at the estimated value of the critical coupling in
         the bulk system $K_{c}=0.535$ and at the maximum of $\frac{\partial <|M|> }{ \partial K}$,
         versus the linear system size $L$. The straight lines are fits to the data. 
         Their slope is $- \beta / \nu$. The obtained values of the ratio of
         exponents $\beta / \nu$ are indicated.}  
\label{f7} 
\end{figure}

The exponent ratio $\beta / \nu$ can be obtained from the finite-size scaling behaviour of $|M|$ 
either at $K=K_{c}$ or at the value of $K$ where the derivative $\partial <|M|> / \partial K$ 
has the maximum. FSS predicts that $|M|$ at these $K$ values should obey the relation 
$M \sim L^{-\beta / \nu}$.  Figure \ref{f7} shows the plots corresponding to this relation.
The straight lines in this figure were obtained using the least-square fitting routine.
The exponent $\beta / \nu$ was determined from the slope of the lines.  We obtained:
$\beta / \nu = 0.141 \pm 0.005$  $(Q=0.74)$ at $K_{c}$ and $\beta / \nu = 0.120 \pm 0.006$
$(Q=0.77)$ at the maximum of  $\partial <|M|> / \partial K$.
\begin{figure}
\setlength{\epsfxsize}{9cm}
\centerline{\mbox{\epsffile{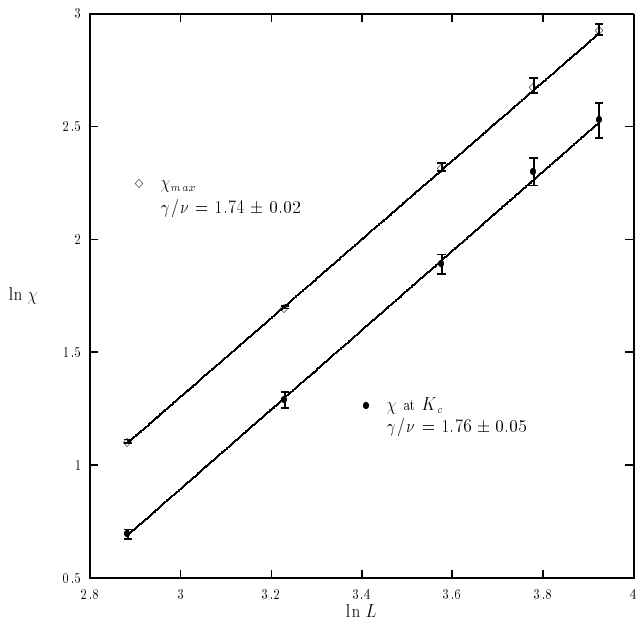}}}

\caption{The plot of the magnetic susceptibility at the estimated value of the critical coupling 
         in the bulk system $K_{c}=0.535$ and at the maximum, 
         versus the linear system size $L$. The straight lines are fits to the data. 
         Their slope is $\gamma / \nu$. The obtained values of the ratio of
         exponents $\gamma / \nu$ are indicated.} 
\label{f8} 
\end{figure}

The exponent ratio $\gamma / \nu$ can be determined from the finite-size scaling behaviour of the 
maximum of the magnetic susceptibility $\chi_{max}$ and of the value $\chi(K_{c})$ of 
the magnetic susceptibility at $K=K_{c}$. According to FSS these quantities are expected to vary 
with system size like $L^{\gamma / \nu}$.
Figure \ref{f8} displays the finite-size scaling behaviour of $\chi_{max}$ and $\chi(K_{c})$.
We estimated the value of the exponent ratio $\gamma / \nu$ from slopes of fitted straight lines.
From values $\chi_{max}$ we obtained $\gamma / \nu = 1.74 \pm 0.02$  $(Q=0.31)$, whereas 
from values $\chi(K_{c})$ we obtained $\gamma / \nu = 1.76 \pm 0.05$  $(Q=0.88)$. 

\section{ Conclusions }

We have studied the critical behaviour of the two-dimensional Ising fluid at a density 
$\rho=0.4$ near the second-order phase transition from the paramagnetic gas to the 
ferromagnetic gas phase.  The multiple histogram technique was applied in order to combine
data obtained from four different MC simulations.  The largest system we studied consisted
of 1024 particles.  The critical exponent $\nu$ was determined
from FSS behaviour of the derivatives $\partial \ln <|M|> / \partial K$ 
and $\partial \ln <M^{2}> / \partial K$.  We obtained $\nu \approx 1.35$. This value is much 
higher than the value $\nu=1$ obtained for the two-dimensional Ising lattice model.
The critical coupling $K_{c}$ in the bulk system was obtained by considering size-dependent
shifting of the maxima of several quantities using the estimated value of the exponent $\nu$.
This value was found to be consistent with the location of the fourth-order magnetization 
cumulant crossing points i.e. with $K_{c} \approx 0.535$. 
The estimated common value of the cumulant at $K=K_{c}$ is $U^{*}=0.618$ what is slightly higher 
than the value $U^{*}=0.611$ obtained in the two-dimensional lattice Ising model. 
All the 
cumulant intersection values given in table I are above the value $0.611$
except the value determined for the pair of systems $[768,1024]$. 
 As the statistical error increases with the system size
and considerable error is expected in the determination of cumulant intersection points 
for pairs of systems with small difference in sizes we have neglected the last three rows of 
table I in estimating
$K_{c}$ and $U^{*}$.
The ratio of critical exponents $\beta / \nu$ and $\gamma / \nu$ we determined from the
FSS behaviour of the magnetization and magnetic susceptibility.   They are consistent with values obtained
in the two-dimensional lattice Ising model.  In annealed diluted magnets the exponents are 
renormalized if the pure specific-heat exponent $\alpha$ is positive and do not change 
if it is negative \cite{r18}.  As for the pure lattice Ising model $\alpha=0$, we expect 
for the annealed diluted Ising models the pure Ising exponents.  Our results
disagree with this prediction and suggests a renormalization of exponents 
which leaves $\beta / \nu$ and $\gamma / \nu$ unchanged and  changes $\nu$. 
We were unable to consider 
large system sizes with good statistics to determine possible corrections to finite-size
scaling. However, the presented self-consistency in the determination of $K_{c}$ using the 
estimated
exponent $\nu$ convinces one that the obtained values of $\nu$ and $K_c$ are correct.

A behaviour similar to ours was reported in studies of the 
two-dimensional  quenched site diluted
Ising model \cite{r19}. These studies suggest 
that the $\nu$ exponent and the Binder cumulant value, $U^{*}$ increases with the degree of 
disorder. These was interpreted as the verification of the weak universality scenario\cite{r20}
also seen to apply to other models\cite{r19}. 
We plan to study our model with different densities to see if
the measured exponents approach the pure Ising values with increasing density.
\begin{table}
\begin{center}
\caption{The values of fourth-order magnetization cumulant \protect{$U_{cross}(N_{1},N_{2})$} and
         the coupling \protect{$K_{cross}(N_{1},N_{2})$} at the intersection point for 
different pairs of systems \protect{$[N_{1},N_{2}]$} having \protect{$N_{1}$} and 
\protect{$N_{2}$} particles.}

\label{t1}
\begin{tabular}{l c c}
 $[N_{1},N_{2}]$    &   $K_{cross}(N_{1},N_{2})$   &  $U_{cross}(N_{1},N_{2})$  \\
 \hline  
 $[128,256]$         &   0.5327                     &  0.6153           \\ 
 $[128,512]$         &   0.5331                     &  0.6158           \\ 
 $[128,768]$         &   0.5364                     &  0.6193          \\ 
 $[128,1024]$        &   0.5358                     &  0.6186          \\  
 $[256,512]$         &   0.5334                     &  0.6164         \\ 
 $[256,768]$         &   0.5380                     &  0.6224          \\ 
 $[256,1024]$        &   0.5368                     &  0.6209          \\ 
 $[512,768]$         &   0.5449                     &  0.6336           \\ 
 $[512,1024]$        &   0.5398                     &  0.6269           \\ 
 $[768,1024]$        &   0.5324                     &  0.6105           \\
\end{tabular}
\end{center}
\end{table} 
\section*{Acknowledgments} 
W.K. thanks the Junta Nacional de Investiga\c c\~ao Cient\'\i fica e Tecnol\'ogica 
in Portugal (JNICT) for the grant under the program PRAXIS XXI, 
and Professor S.K.Mendiratta for his kind hospitality at 
the University of Aveiro. A. L Ferreira thanks JNICT for support under PRAXIS2/2.1/FIS/299/94.
The references \cite{r19} and \cite{r20} were pointed out to us by M. A. Santos 
after the first version of the paper was completed.

                           
%

\end{multicols}

\end{document}